\documentclass{appolb}%
\usepackage{epsfig}
\usepackage{rotating}
\usepackage{hyperref}
\usepackage{comment,color}
\usepackage{braket}
\usepackage{amssymb,amsmath,bm}
\usepackage{setspace,lipsum}
\usepackage{amsmath}
\usepackage{amsfonts}
\usepackage{amssymb}
\usepackage{graphicx}%
\providecommand{\U}[1]{\protect\rule{.1in}{.1in}}
\begin{document}

\title{ QZE and IZE in a simple approach and the neutron decay}
\author{Francesco~Giacosa
\address{Institute of Physics, Jan Kochanowski University, Ul. Uniwersytecka 7, 25-406 Kielce, Poland}
}
\maketitle

\begin{abstract}
We discuss a simple and analytically solvable measurement model which
describes the famous Quantum Zeno Effect (QZE) and Inverse Zeno Effect (IZE),
that correspond to the slow down and to the increase of the decay rate caused
by measurements (or, more in general, by the interaction of an unstable state
with the detector and the environment). Within this model one can understand
quite general features of the QZE and IZE: by considering an unstable quantum
state, such as an unstable particle, whose decay width as function of energy
is $\Gamma(\omega)=g^{2}\omega^{\alpha},$ then -under quite general
assumptions- the QZE occurs for $\alpha\in(0,1)$, while the IZE for $\alpha
\in(-\infty,0)\cup(1,\infty).$ This result is valid also for more realistic
measurement models than the one described in this work. We then apply these
considerations to the decay of the neutron, for which $\alpha=5.$ Hence, the
realization of the IZE for the neutron decay (and for the majority of weak
decays) is in principle possible. Indeed, trap experiments find a lifetime
that is $8.7\pm2.1$ s shorter than beam experiments, suggesting that the IZE
could have taken place.

\end{abstract}


\section{Introduction}

The quantum Zeno effect (QZE) and the inverse Zeno effect (IZE) are the slow
down and the increase of the decay rate of an unstable state (or particle)
when it is `measured' often enough, see e.g. Refs.
\cite{dega,sudarshan,kk,fp,nakazato,fptopicalreview} and Ref. \cite{koshino}
for a review. Both the QZE and the IZE are a consequence of the fact that the
decay law is not exactly exponential
\cite{khalfin,ghirardi,mercouris,raczynska,zenoqft,duecan}. The experimental
verification of nonexponential decays both at short and long times can be
found in Refs. \cite{raizen1,rothe}, while the QZE and IZE on a genuine
unstable quantum system (tunnelling through an optical potential) is described
in\ Ref. \cite{raizen2}.

Basically, each system can undergo the QZE if probed at short enough time
intervals. Here, with `probed' we do not necessarily mean a standard textbook
measurement, but also a sufficiently strong interaction of the system with the
environment can lead to a decoherence which is for all practical purposes
analogous to a measurement. As discussed in Refs. \cite{kk,fp,fptopicalreview}%
, the IZE can be even easier to be realized than the QZE if some conditions
are met, most importantly a strong -but not too strong- system-environment interaction.

In this work, we first briefly review in Sec. 2 the main features of the QZE
and IZE: the key element is the so-called response function, which models the
environment-system interaction. If the system is -in average- probed at a
certain given rate, even if the underlying decay law is not an exponential,
one stills measure an exponential decay law, whose decay width (the inverse of
the lifetime) is however different from the `on-shell' or bare value obtained
in the limit in which the unstable state weakly couples to the environment
(for instance, by doing a single collapse measurement after a sufficiently
long time). The effective or measured decay width emerges as the average of
the decay width as function of the energy convoluted with the previously
mentioned response function.

Next, in\ Sec. 3, we present a simple model for the response function which
allows to present in a clear way under which conditions the QZE and under
which conditions the IZE are realized. Even if this model may be regarded as
`\textit{too simple}' to be realistic, one can understand also results that go
beyond the specific employed form. Namely, one can understand why the IZE is
actually even more general than the QZE. By denoting the decay width function
as $\Gamma(\omega)$ and assuming that in the energy range of interest
$\Gamma(\omega)=g^{2}\omega^{\alpha},$ then we show that -under quite broad
assumptions- the QZE occurs for $\alpha\in(0,1)$, while the IZE for the much
broader range $\alpha\in(-\infty,0)\cup(1,\infty).$

In particular, the IZE is expected to take place in the case of weak decays,
since $\alpha>1$. As a specific applications of our considerations, we present
in Sec. 4 the example of the decay of the neutron, for which $\alpha=5$. At
present an anomaly exists \cite{wietfeldt}: beam experiments which measure the
emitted protons find the lifetime $\tau_{\mathrm{n}}^{\text{beam}}%
=888.1\pm2.0$ s, while trap or cavity experiments which monitor the surviving
neutrons deliver the result $\tau_{\mathrm{n}}^{\text{trap}}=879.37\pm0.58$ s.
As recently proposed in\ Ref. \cite{izen}, the possibility that the mismatch
is due to the IZE realized in trap experiments is discussed.

Finally, in\ Sec. 5 we present our conclusions.

\section{QZE and IZE: general discussion}

In this section we summarize the results of Refs.
\cite{kk,fp,nakazato,fptopicalreview}, where a theoretical approach for the
description of the measurement has been put forward. A decay process of an
unstable state (or particle) called `$n$' is described by the decay function
$\Gamma(\omega)$. The energy $\omega$ reads
\begin{equation}
\omega=m-\sum_{j=1}^{N}m_{j}\text{ ,}%
\end{equation}
where $m_{j}$ are the `masses' (or energies) of the $N$ decay products of the
state $n$ and $m$ is the `running mass' of the state. The quantity $\omega$
(and so $m$) can vary, since the mass of an unstable state is not fixed.
Moreover, $\omega\geq0$, since the running mass cannot be smaller than the sum
of the masses in the final state. The on-shell value is obtained for
\begin{equation}
\omega_{\text{on-shell}}\equiv\omega_{n}=m_{n}-\sum_{j=1}^{N}m_{j}\text{ ,}%
\end{equation}
where
\begin{equation}
m_{n}=m_{\text{on-shell}}\text{ .}%
\end{equation}
The `on-shell' decay width is given by%
\begin{equation}
\Gamma_{n}=\Gamma(\omega_{n})=\Gamma_{\text{on-shell}}=\frac{1}{\left\langle
t_{n}\right\rangle }\text{ ,} \label{osw}%
\end{equation}
where $\left\langle t_{n}\right\rangle $ is the mean lifetime of the unstable
state $n.$

The form of the function $\Gamma(\omega)$ can be evaluated in the framework of
the given model/approach. One possibility goes through the so-called Lee model
\cite{lee} (see also Refs. \cite{duecan,delta,x3872,lonigro,leerev} and refs.
therein) or within a certain given quantum field theoretical approach, e.g.
Refs. \cite{peskin,greiner} (for the link of QFT to nonexponential decays, see
also Refs. \cite{zenoqft,duecan,lupo}).

A general result is that, in presence of a series of measurements and/or
interactions of the system with the environment, the effective measured decay
width may change according to the weighted average:
\begin{equation}
\Gamma^{\text{measured}}(\tau)=\int_{0}^{\infty}f(\tau,\omega)\Gamma
(\omega)d\omega\text{ ,} \label{gamman1}%
\end{equation}
where the parameter $\tau=\lambda^{-1}$ (with $\lambda$ being the
corresponding rate) describes how strong is the coupling of the environment
with the system: large $\tau$ (small $\lambda)$ means weak coupling (in which
one should recover the on-shell decay width of Eq. (\ref{osw})), while small
$\tau$ (large $\lambda)$ implies a strong coupling, in which deviations from
the on-shell width are expected.

The function $f(\tau,\omega)$ can be regarded as the `response function' of
the environment/detector on the quantum system. Its form is generally peaked
and symmetric w.r.t. $\omega_{n}$, but the details depend on the
system-environment-detector interaction(s). Nevertheless, three general
constraints are:
\begin{equation}
\text{(i) }\int_{0}^{\infty}f(\tau,\omega)d\omega=1\text{ ; } \label{cond1}%
\end{equation}

\begin{equation}
\text{(ii) }f(\tau\rightarrow\infty,\omega)=\delta(\omega-\omega_{n})\text{ ;}
\label{cond2}%
\end{equation}

\begin{equation}
\text{(iii) }f(\tau\rightarrow0,\omega)=\text{small constant .} \label{cond3}%
\end{equation}

The first condition in Eq. (\ref{cond1}) guarantees the normalization. As a
consequence, in the Breit-Wigner limit, in which $\Gamma(\omega)=\Gamma
_{\text{BW }}$ is a simple constant and no deviation from the exponential
decay occurs \cite{ww}, one has
\begin{equation}
\Gamma^{\text{measured}}(\tau)=\int_{0}^{\infty}f(\tau,\omega)\Gamma
(\omega)d\omega=\Gamma_{\text{BW }}\int_{0}^{\infty}f(\tau,\omega
)d\omega\,=\Gamma_{\text{BW }}\text{ ,} \label{gammameas}%
\end{equation}
for each measurement function $f(\tau,\omega).$ Then, as expected, neither QZE
nor the IZE can take place. This case is however unphysical, since a constant
decay width and the corresponding Breit-Wigner distribution are only an
approximation (which in many cases is so good that it is hard to see any difference).

The second condition in Eq. (\ref{cond2}) assures that, if the system is
undisturbed, one obtains the `on-shell' free decay width
\begin{equation}
\Gamma^{\text{measured}}(\tau\rightarrow\infty)=\Gamma_{n}=\Gamma
_{\text{on-shell}}\text{ .}%
\end{equation}

Finally, the third condition in Eq. (\ref{cond3}) implies that, for $\tau$
very small, $f(\tau\rightarrow0,\omega)$ is a (small) constant, hence
\begin{equation}
\Gamma^{\text{measured}}(\tau\rightarrow0)=(\text{small constant})\int
_{0}^{\infty}\Gamma(\omega)d\omega\,\rightarrow0\text{ }, \label{qze}%
\end{equation}
assuming the convergence of $\int_{0}^{\infty}\Gamma(\omega)d\omega\,$: this
is the famous QZE mentioned above.

The functional form of $f(\tau,\omega)$ depends on which type of measurement
is performed. Two famous forms were considered in\ Refs.
\cite{kk,fptopicalreview}. For the case of instantaneous ideal bang-bang
measurements at time intervals $0,\tau,$ $2\tau,...$ one gets
\begin{equation}
f(\tau,\omega)=\frac{\tau}{2\pi}\frac{\sin^{2}\left[  \tau\left(
\omega-\omega_{n}\right)  /2\right]  }{[\tau\left(  \omega-\omega_{n}\right)
/2]^{2}}\text{ }. \label{f1}%
\end{equation}
In the case of a continuous measurement (in the form e.g. of a continuous
detector-system interaction, see Refs. \cite{kk,fp} for details; for the
general concept of a continuous monitoring, see also Ref.
\cite{schulman,streed,giacosapra}) one gets:
\begin{equation}
f(\tau,\omega)=\frac{1}{\pi\tau}\frac{1}{\left(  \omega-\omega_{n}\right)
^{2}+\tau^{-2}}\text{ }. \label{f2}%
\end{equation}
More in general, the response function is not solely caused by measurements.
The time-scale $\tau$ may be regarded as the dephasing/ decoherence time for
the whole environment-object-detector system. Actually, in various physical
examples, the value of $\tau$ determined by the environment is more efficient
than the actual measurements performed by a detector \cite{omnes}.

One may also note that the very convergence of Eq. (\ref{gammameas}) is not
necessarily guaranteed. This is why in various applications one needs to
further restrict the off-shellness of the unstable state to a certain range,
upon replacing $\int_{0}^{\infty}d\omega\lbrack...]\rightarrow\int_{\omega
_{n}-\Delta E}^{\omega_{n}+\Delta E}d\omega\lbrack...].\ $Moreover, also the
normalization (i) of Eq. (\ref{cond1}) is not fulfilled for the functions in
Eqs. (\ref{f1}) and (\ref{f2}) (even if numerically very well realized).

In the next section, we describe a simple model which fulfills all conditions
(i), (ii), and (iii) exactly and -in addition- guarantees always the
convergence of $\Gamma^{\text{measured}}(\tau).$

\section{QZE and IZE: a simple model}

Here, as a concrete and simple model we introduce the following rectangular
response function:
\begin{equation}
f_{\text{rect}}(\tau=\lambda^{-1},\omega)=N_{\lambda}\theta(\omega
)\theta(\lambda^{2}-(\omega-\omega_{n})^{2})\text{ }. \label{rect}%
\end{equation}
The constraint $N_{\lambda}$ is necessary to guarantee condition (i) of Eq.
(\ref{cond1}):%
\begin{equation}
N_{\lambda}=\left\{
\begin{array}
[c]{c}%
\frac{1}{2\lambda}\text{ for }\omega_{n}-\lambda>0\\
\frac{1}{\omega_{n}+\lambda}=\frac{1}{\omega_{C}}\text{for }\omega_{n}%
-\lambda>0
\end{array}
\right.  \text{ ,}%
\end{equation}
where the upper limit $\omega_{C}=\omega_{n}+\lambda$ has been introduced.
Note, for $\omega_{n}-\lambda>0$ the function takes the form
\begin{equation}
f_{\text{rect}}(\tau=\lambda^{-1},\omega)=\left\{
\begin{array}
[c]{c}%
0\text{ for }\left\vert \omega-\omega_{n}\right\vert >\lambda\\
\frac{1}{2\lambda}\text{for }\left\vert \omega-\omega_{n}\right\vert
\leq\lambda
\end{array}
\right.
\end{equation}
It then follows that for $\lambda\rightarrow0$ (that is, $\tau\rightarrow
\infty)$ the function $f_{\text{rect}}(\tau=\lambda^{-1},\omega)$ is a
possible representation of the $\delta$-function:
\begin{equation}
f_{\text{rect}}(\tau\rightarrow\infty,\omega)=\delta(\omega-\omega_{n})\text{
,}%
\end{equation}
hence the condition (ii) of Eq. (\ref{cond2}) is also guaranteed:
\begin{equation}
\Gamma^{\text{measured}}(\tau\rightarrow\infty)=\int_{0}^{\infty
}f_{\text{rect}}(\tau\rightarrow\infty,\omega)\Gamma(\omega)d\omega
=\Gamma(\omega_{n})=\Gamma_{\text{on-shell}}\text{ .}%
\end{equation}

Next, let us consider the limit $\lambda\rightarrow\infty.$ It is then clear
that
\begin{equation}
f_{\text{rect}}(\tau=\lambda^{-1},\omega)=\frac{1}{\omega_{n}+\lambda}%
\theta(\omega)\theta(\omega_{n}+\lambda-\omega)\overset{\lambda\gg\omega_{n}%
}{\simeq}\frac{1}{\lambda}\theta(\omega)\theta(\lambda-\omega)\text{ }.
\end{equation}
Ergo, the\ QZE is easily realized (condition (iii) of Eq. (\ref{cond3})):
\begin{equation}
\Gamma^{\text{measured}}(\tau)=\int_{0}^{\infty}f_{\text{rect}}(\tau
\rightarrow0,\omega)\Gamma(\omega)d\omega=\lim_{\lambda\rightarrow\infty}%
\frac{1}{\lambda}\int_{0}^{\lambda}\Gamma(\omega)d\omega=0\text{ ,}
\label{qze2}%
\end{equation}
as long as $\int_{0}^{\infty}\Gamma(\omega)d\omega$ is finite (as it must be
in each physical case).

Thus, all the conditions are fulfilled and -in addition- the response function
cuts abruptly energies outside a certain range and is constant within a given
range. This is different from Eqs. (\ref{f1}) and (\ref{f2}): the question if
the response function in\ Eq. (\ref{rect}) can be -at least in some cases-
partially realistic is hard to answer. Yet, as we shall see below, it is
useful to show quite general properties.

Next, let us consider the case in which $\omega_{n}-\lambda>0$ and assume
that, within the range ($\omega_{n}-\lambda,\omega_{n}-\lambda)$ we can
approximate the decay function as
\begin{equation}
\Gamma(\omega)=g^{2}\omega^{\alpha}\text{ for }\omega\in(\omega_{n}%
-\lambda,\omega_{n}-\lambda)\text{ .} \label{norm}%
\end{equation}
This is only an approximation, but as long as $\lambda$ is small enough one
may consider $g^{2}\omega^{\alpha}$ as the dominant contribution. Yet, even in
the case in which this approximation is not possible, one can always consider
$\Gamma(\omega)$ as a polynomial function, hence one can easily generalize the
argument that we are about to present. Indeed, if $\lambda$ is very large,
also such an approximation would break down. A typical expression that would
include a form factor is given by
\begin{equation}
\Gamma(\omega)=g^{2}\omega^{\alpha}e^{-(\omega-\omega_{n})^{2}/\Lambda^{2}%
}\text{ ,} \label{norm2}%
\end{equation}
which guarantees the necessary convergence to guarantee and the realization of
QZE, See Eqs. (\ref{qze}) and (\ref{qze2}), if $\lambda$ is large enough.

We come back to the approximation of Eq. (\ref{norm}). The integral can be
solved exactly:
\begin{align}
\Gamma^{\text{measured}}(\tau)  &  =\frac{1}{2\lambda}\int_{\omega_{n}%
-\lambda}^{\omega_{n}+\lambda}\Gamma(\omega)d\omega=\frac{g^{2}}{2\lambda
}\frac{1}{\alpha+1}\left(  \omega^{\alpha+1}\right)  _{\omega_{n}-\lambda
}^{\omega_{n}+\lambda}\\
&  =\frac{g^{2}}{2\lambda}\frac{1}{\alpha+1}\left(  \left(  \omega_{n}%
+\lambda\right)  ^{\alpha+1}-\left(  \omega_{n}-\lambda\right)  ^{\alpha
+1}\right) \\
&  =\frac{g^{2}}{2\lambda}\frac{\omega_{n}^{\alpha+1}}{\alpha+1}\left(
\left(  1+x\right)  ^{\alpha+1}-\left(  1-x\right)  ^{\alpha+1}\right)
\end{align}
where the ratio
\begin{equation}
x=\frac{\lambda}{\omega_{n}}%
\end{equation}
has been introduced. The number $x$ is expected to be safely smaller than
unity. Next, let us consider the following Taylor expansion up to third order:%
\begin{equation}
(1+x)^{\alpha+1}=1+(\alpha+1)x+\frac{1}{2}(\alpha+1)\alpha x^{2}+\frac{1}%
{3!}(\alpha+1)\alpha(\alpha-1)x^{3}+...
\end{equation}
Note, going up to $x^{3}$ is necessary for our purposes. By plugging in:
\begin{align}
\Gamma^{\text{measured}}(\tau)  &  =\frac{g^{2}}{2\lambda}\frac{\omega
_{n}^{\alpha+1}}{\alpha+1}\left[  1+(\alpha+1)x+\frac{1}{2}(\alpha+1)\alpha
x^{2}+\frac{1}{3!}(\alpha+1)\alpha(\alpha-1)x^{3}+...\right. \nonumber\\
&  \left.  -\left(  1-(\alpha+1)x+\frac{1}{2}(\alpha+1)\alpha x^{2}-\frac
{1}{3!}(\alpha+1)\alpha(\alpha-1)x^{3}\right)  \right]  \text{ ,}%
\end{align}
then%
\begin{align}
\Gamma^{\text{measured}}(\tau)  &  =\frac{g^{2}}{2\lambda}\frac{\omega
_{n}^{\alpha+1}}{\alpha+1}\left[  2(\alpha+1)x+\frac{2}{3!}(\alpha
+1)\alpha(\alpha-1)x^{3}+...\right] \nonumber\\
&  =\frac{g^{2}}{2\lambda}\frac{\omega_{n}^{\alpha}\omega_{n}^{\alpha}}%
{\alpha+1}\left[  2(\alpha+1)\frac{\lambda}{\omega_{n}^{\alpha}}+\frac{2}%
{3!}(\alpha+1)\alpha(\alpha-1)\left(  \frac{\lambda}{\omega_{n}}\right)
^{3}+...\right] \\
&  =g^{2}\omega_{n}^{\alpha}\left[  1+\frac{1}{3!}\alpha(\alpha-1)\left(
\frac{\lambda}{\omega_{n}}\right)  ^{2}+...\right]  \text{ .}%
\end{align}
We then find the following result, which can be regarded as the main
achievement of the present work:
\begin{equation}
\Gamma^{\text{measured}}(\tau)=\Gamma_{n}\left[  1+\frac{\alpha(\alpha-1)}%
{6}\frac{\lambda^{2}}{\omega_{n}^{2}}+\mathcal{O}\left(  \frac{\lambda^{4}%
}{\omega_{n}^{4}}\right)  \right]  \text{ .}%
\end{equation}
One sees that the result depends on $\alpha$ and in particular on the sign of
the quantity $\alpha(\alpha-1).$ We have:
\begin{equation}
\text{QZE: }\Gamma^{\text{measured}}(\tau)<\Gamma_{n}\text{ if }%
0<\alpha<1\text{ ;} \label{aqze}%
\end{equation}
the well-known QZE is realized. We recall that the QZE is anyhow realized if
$\tau$ is small enough ($\lambda$ large enough, see Eqw. (\ref{qze}) and
(\ref{qze2})), but it can be also realized for a relatively large value of
$\tau$ if the condition $0<\alpha<1$ is met.

Next, the IZE takes place for:%
\begin{equation}
\text{IZE: }\Gamma^{\text{measured}}(\tau)>\Gamma_{n}\text{ if }\alpha<0\text{
or }\alpha>1\text{ .} \label{aize}%
\end{equation}
Thus, one can see that the IZE is actually easier to obtain than the QZE (of
course, for sufficiently small (but not too small) $\tau=\lambda^{-1}$). In
most physical cases, indeed $\alpha>1$ is realized.

In between one has%
\begin{equation}
\Gamma^{\text{measured}}(\tau)=\Gamma_{n}\text{ for }\alpha=0\text{ and
}\alpha=1. \label{a0}%
\end{equation}
This result is expected for $\alpha=0$ (this is the Breit-Wigner limit) but,
quite interestingly, holds also for $\alpha=1$, when $\Gamma(\omega
)=g^{2}\omega$ is linear.

This result can be actually extended to any symmetric response function:
\begin{equation}
f(\tau,\omega)=\sum_{k}c_{k}f_{\text{rect}}(\tau_{k}=\lambda_{k}^{-1}%
,\omega)\text{ } \label{sumck}%
\end{equation}
where all $c_{k}$ are positive functions of $\tau$ and are such that $\sum
_{k}c_{k}=1.$ For instance, the functions in\ Eqs. (\ref{f1}) and (\ref{f2})
can be re-expressed in this way. It follows that%
\begin{equation}
\Gamma^{\text{measured}}=\Gamma_{n}\left[  1+\sum_{k}c_{k}\frac{\alpha
(\alpha-1)}{6}\frac{\lambda^{2}}{\omega_{n}^{2}}+\mathcal{O}\left(
\frac{\lambda^{4}}{\omega_{n}^{4}}\right)  \right]  \text{ ,}%
\end{equation}
hence the final results of Eqs. (\ref{aqze}), (\ref{aize}) and (\ref{a0}) are
still valid for a generic response function that fulfills Eq. (\ref{sumck}).

Another interesting case is obtained when two (or more) terms are present (for
instance as in the case of different decay channels, an interesting topic in
non-exponential decay \cite{duecan,capturing}):
\begin{equation}
\Gamma(\omega)=g_{1}^{2}\omega^{\alpha}+g_{2}^{2}\omega^{\beta},
\end{equation}
out of which%
\begin{equation}
\Gamma^{\text{measured}}(\tau)=\Gamma_{n}^{(1)}\left[  1+\frac{\alpha
(\alpha-1)}{6}\frac{\lambda^{2}}{\omega_{n}^{2}}\right]  +\Gamma_{n}%
^{(2)}\left[  1+\frac{\beta(\beta-1)}{6}\frac{\lambda^{2}}{\omega_{n}^{2}%
}\right]  .
\end{equation}
It is then clear that if both $\alpha$ and $\beta$ $\in(0,1)$ the QZE is
realized, while otherwise the IZE takes place. Yet, if $\alpha$ $\in(0,1)$ and
$\beta$ doesn't (or vice-versa), then there are two conflicting phenomena and
no general statement can be made: the precise values of the coupling constants
is necessary to assess if the decay width has decreased or increased.

Next, we consider the case in which $\omega_{n}$ is close enough to $0$ (the
lowest possible energy) such that $\lambda>\omega_{n}$. In this case one has
\begin{align}
\Gamma^{\text{measured}}(\tau)  &  =\int_{0}^{\infty}f_{\text{rect}}%
(\tau,\omega)\Gamma(\omega)d\omega=\frac{1}{\omega_{C}}\int_{0}^{\omega_{C}%
}\Gamma(\omega)d\omega\\
&  =\frac{1}{\omega_{C}}g^{2}\frac{\omega_{C}^{\alpha+1}}{\alpha+1}=\Gamma
_{n}\frac{1}{\alpha+1}\frac{\omega_{C}^{\alpha}}{\omega_{n}^{\alpha}}\\
&  =\Gamma_{n}\left[  1+\frac{\left(  \omega_{C}/\omega_{n}\right)  ^{\alpha
}-\left(  \alpha+1\right)  }{\alpha+1}\right]  \text{ ,}%
\end{align}
where $\omega_{C}=\omega_{n}+\lambda\geq2\omega_{n}.$ Then, one has:
\begin{equation}
\text{IZE for }\alpha<0\text{ and for }\alpha>\alpha_{0}>0
\end{equation}
with%
\begin{equation}
\left(  \omega_{C}/\omega_{n}\right)  ^{\alpha_{0}}=\alpha_{0}+1.
\end{equation}
Since $\omega_{C}/\omega_{n}\geq2,$ it turns out that $\alpha_{0}<1$:
\textit{the range for the IZE increases even further}.

The QZE is confined to the interval
\begin{equation}
\text{QZE for }0<\alpha<\alpha_{0}<1,
\end{equation}
thus the corresponding range decreased. Finally:%
\begin{equation}
\Gamma^{\text{measured}}(\tau)=\Gamma_{n}\text{ for }\alpha=0\text{ and
}\alpha=\alpha_{0}<1\text{ }.
\end{equation}

\section{The neutron decay}

Let us finally discuss two physical examples. First, we discuss the case of
the neutron, since as mentioned in the introduction it is particularly
interesting in view of a persisting anomaly. The neutron decay is a weak decay
whose decay width function has the from
\begin{equation}
\Gamma(\omega)=g_{n}^{2}\omega^{5}\,
\end{equation}
$\,$(this is actually the leading term, for the full formula see e.g. Ref.
\cite{greiner}), thus $\alpha=5$: the IZE is possible. The on-shell values are
$\omega_{\text{onshell}}=m_{n}-m_{p}-m_{e}=0.782333$ MeV and $\Gamma
_{\text{onshell}}=g_{n}^{2}\omega_{\text{onshell}}^{5}=\hslash/\tau_{\text{n}%
}^{\text{beam}}=\hslash/888.1$ sec$^{-1}=7.41146\cdot10^{-25}$ MeV (implying
$g_{n}=1.59028\cdot10^{-12}$ MeV$^{-2}$). [Note, the anyhow small errors are
neglected here]. Of course, the function $\Gamma(\omega)$ cannot rise
indefinitely. An expression of the type as in\ Eq. (\ref{norm}) is expected to
hold. To be more precise, the following behavior for the neutron is
realistic:
\begin{equation}
\Gamma(\omega)\propto\left\{
\begin{array}
[c]{c}%
\omega^{5}\text{ for }\omega\lesssim M_{W}\\
\omega\text{ for }M_{W}\lesssim\omega\lesssim M_{X}\\
\omega e^{-\omega/\Lambda}\text{ for }\omega\gtrsim M_{X}%
\end{array}
\right.  .
\end{equation}
where $M_{X}$ is some large scale, $X=GUT$ or \ $M_{Planck.}$ Also $\Lambda$
is some large number of the order of $M_{X}.$

Ergo, for $\lambda$ up to $M_{W}$ we are basically in the IZE regime: that
means that in practice only the IZE is realistic for the neutron. For even
larger $\lambda$ the contribution $\propto\omega$ enters \cite{lupo} and only
for(an irrealistic) $\lambda$ larger than $M_{X}$ the decreasing of
$\Gamma^{\text{measured}}(\tau)$ would start to be visible.

Let us now discuss the IZE for ongoing experiments. In beam experiments, the
value of $\tau$ was estimated in\ Ref. \cite{izen} to be quite large, thus
$\lambda$ turns out to to be very small, sizably smaller than $10^{-6}$ MeV.
Hence, the IZE is very small; to a very good extent:
\begin{equation}
\Gamma^{\text{measured-beam}}(\tau_{\text{beam}})\simeq\Gamma_{\text{on-shell}%
}\text{ }.
\end{equation}
On the other hand, for trap experiments, neutrons are kept in a very cold trap
and they are constantly monitored by the environment. Together with the high
degree of correlation in the wave function, it was proposed in Ref.
\cite{izen} that $\tau$ can be sizably smaller, hence $\lambda$ could be
sufficiently large for a sizable IZE. Using the model explained in\ Sec. 3,
for the value $\lambda=0.0424$ MeV one obtains $\Gamma^{\text{measured}}%
(\tau=\lambda^{-1})/\Gamma_{\text{onshell}}=1.0098$. Namely, in this way one
could understand why the trap experiment find a larger decay. We also refer to
\cite{giacosaacta1} in which this topic is discussed by using both response
functions presented in\ Sec. 2. The results were found to be very similar.
More in general, even if the present neutron anomaly is due to some systematic
effects, one may still speculate that the IZE can be realized in future
experimental setups.

As a second example, we also mention the decay of the muon, for which also
$\alpha=5$. Here, $\omega_{\text{on-shell}}\simeq m_{\mu}=105.658$ MeV and
$\Gamma_{\mu}=\hslash/(2.19698\cdot10^{-6}$ sec$)$ \cite{pdg}. In order to get
an increase of $1\%$ on this value, one would need $\lambda=5.787$ MeV, which
is quite large. It seems therefore hard to measure the IZE in experiments
involving muons.

\section{Conclusions}

In this work we have introduced a simple measurement model, based on a
rectangular response function, that allows to understand under which
conditions the QZE and the IZE are realized. We have found that for realistic
measurements the IZE is actually favoured w.r.t. the QZE. Namely, when the
decay widths scales as $\omega^{\alpha}$ with $\alpha>1$ or $\alpha<0$ the IZE
takes place, while the QZE is possible only for $0<\alpha<1.$ The latter
interval is further reduced if the on-shell energy $\omega_{\text{on-shell}}$
is close to zero (the left-energy threshold).

We have then applied these ideas to the decay of the neutron. In recent works
an experimental anomaly between trap and beam experiments has been found
\cite{wietfeldt}: the lifetime measured in trap experiments -in which neutrons
are monitored- is shorter than then one in beam experiments, where protons are
counted. This mismatch has been interpreted in\ Refs. \cite{fornal,berezhiani}
as the effect of a \textquotedblleft\textit{beyond standard model}%
\textquotedblright\ (BSM) invisible dark decay of the neutron that is
undetected in the beam method. This idea has been criticized in Ref.
\cite{baym}, according to which a dark neutron would undermine the stability
of neutron stars, as well as in\ Refs. \cite{dubbers,czarnecki}, where it is
shown that the present standard model result coincides with the beam method.
The conclusion would be that there is some unknown systematic error that
affects the beam experimental setup.

In our approach, there is no need to use BSM physics: the IZE takes place for
the case of neutron decays in traps, hence the shorter lifetime w.r.t. the
beam result. Yet, if the standard model result is correct
\cite{dubbers,czarnecki} and the present anomaly is just an experimental
artifact, there is also no need for an\ IZE in trap experiments. Nevertheless,
our study shows that, for neutron decays, the IZE is not far from reach, hence
it could be measured in future experiments dealing with cold neutrons.

\bigskip

\textbf{Acknowledgments:} The author thanks G. Pagliara for collaboration
leading to Ref. \cite{izen}. F.G. acknowledges support from the Polish
National Science Centre (NCN) through the OPUS no. 2018/29/B/ST2/02576.

\end{document}